\documentclass{ws-ijmpa}
\usepackage[super,compress]{cite}
\usepackage{graphicx}

\usepackage{isotope}
\usepackage{physics}

\begin{document}
\markboth{Sergei~P.~Maydanyuk and Gyorgy~Wolf}{Structure of nuclei in dileptons production in proton-nucleus scattering}

%
\catchline{}{}{}{}{}
%
\title{Structure of nuclei in dileptons production in proton-nucleus scattering} 

\author{Sergei~P.~Maydanyuk$^{(1,2,3)}$\footnote{sergei.maydanyuk@impcas.ac.cn},
Gyorgy~Wolf$^{(2)}$\footnote{wolf.gyorgy@wigner.hu}
}

\address{$^{(1)}$Institute of Modern Physics, Chinese Academy of Sciences, Lanzhou, 730000, China\\
$^{(2)}$Wigner Research Centre for Physics, Budapest, 1121, Hungary\\
$^{(3)}$Institute for Nuclear Research, National Academy of Sciences of Ukraine, Kyiv, 
Ukraine}
\maketitle



\begin{abstract}
We investigate production of electron-positron pairs (dileptons) in the scattering of protons off nuclei.
Focus is directed on clarifying role of nuclear interactions
which make basis in mechanisms of scattering process and structure of nuclei.
For that, we constructed a new model of production of dileptons, where scattering of nuclei and their structure are described on the basis of quantum mechanics.
Cross sections of dilepton production are calculated in the scattering of protons on \isotope[9]{B} at energy of proton beam 
of 2.1~GeV.
Tendency of the calculated full spectrum is in good agreement with experimental data of DLS Collaboration.
%
Contribution of incoherent processes has a leading role in production of dileptons in comparison with coherent one.
%
In our model calculated cross section of dileptons is sensitive to nuclear part of potential of interactions between proton and nucleus in scattering.
%
Influence of structure of nucleus on calculations of cross sections of production of dileptons is essential.
This result provides save basis for new opportunity to extract new information about structure of nuclei, nuclear part of potential from experimental data of dileptons
at studied energies of proton beam.
%
\end{abstract}


\section{Introduction
\label{sec.introduction}}

Heavy-ion collisions in the energy regime from roughly 20 MeV/nucleon to 2 GeV/nucleon offer good
possibility to investigate dense nuclear matter
\cite{
Stocker.Greiner.1986.PhysRep,
Bertsch.DasGupta.1988.PhysRep}.
Compressed dense matter in heavy ion collisions is hot topic studied last few decades \cite{Wolf.1993.PPNP}.
Particles produced
as mesonic and electromagnetic probes provide information about compressive phase of nuclear matter
which can be formed during collision of nuclei
\cite{Mosel.1991.AnnRevNuclPartSci}.
$e^{+} e^{-}$ pairs (dileptons), $\pi$, $\rho$, $\eta$-mesons are
produced in this process%
~\cite{Salabura.2021.PPNP}.

Properties of dilepton production in heavy-ion collisions
at bombarding energies from 1 to 2 GeV/A
were successfully studied by the transport model of the Boltzmann~Uehling-Uhlenbeck (BUU)
type~\cite{Bertsch.DasGupta.1988.PhysRep,Cassing.1990.PhysRep,Wolf.1990.NPA,Bleicher_Bratkovskaya.2022.ProgPartNuclPhys}.
Dynamical evolution of the nucleus-nucleus collision,
nucleon resonances in nuclei,
the electromagnetic form factor of hadrons,
medium properties,
effects of pion and $\Delta$ self-energies on the pion dynamics,
etc.
were well described by such an approach.

As it is well known, understanding 
of nuclear interactions in the studied nuclear process starts when number of nucleons is three or more,
or it needs to include clusters models~\cite{kn:Minn_pot1,1991PhRvC..44.1695L,2015NuPhA.941..121L}.
Study of production of dileptons in proton-nucleus reaction when nucleus is considered as one static medium fragment is not enough,
to obtain realistic description of nuclear process.
So, study of proton-nucleus scattering on the many-nucleon basis is motivated.
Such formalism has obtained success, constructed on the quantum mechanics basis \cite{Maydanyuk_Vasilevsky.2023.PRC,Shaulskyi_Maydanyuk_Vasilevsky.2024.PRC}.
Many purely quantum phenomena such as tunneling for low energies, resonant and potential scatterings, other mechanisms of scattering, formation of bound states in system of nucleons (giving atomic nuclei, fusion, etc.) obtain good description 
for study with desirable accuracy.

The simple understanding on many-nucleon effects in production of dileptons can be obtained from coherent and incoherent processes.
Such effects were studied in emission of bremsstrahlung photons in nuclear reactions within unified model with good accuracy~\cite{Maydanyuk.2023.PRC.delta}.
So, it is natural task to extend this model also for description of production of dileptons in nuclear collisions.
Some basis of this model was given in Ref.~\cite{Maydanyuk_Wolf.2024.arxiv}, but only few physical effects were studied on its basis.
Actually role of nuclear interactions (used in basis of mechanisms of scattering, structure of nuclei) was not analyzed yet.
This is a main task of this paper.

Here we will not consider channels related with resonances, meson decays,
Dalitz decay~\cite{Zetenyi_Wolf.2012.PRC,Wolf.1993.PPNP}.
The simplest nuclear reaction with incoherent processes in dilepton production is the proton nucleus scattering~\cite{Maydanyuk.2023.PRC.delta},
so, we will study it in this paper.
In the 1--5 GeV/nucleon beam energy range the dileptons production was studied by
the DiLepton Spectrometer (DLS) at LBL \cite{Wilson.DSL-collab.1998.PRC,Porter.DSL-collab.1997.PRL} and
the High Acceptance Di-Electron Spectrometer (HADES) at
GSI~\cite{Agakishiev.HADES-collab.2012.plb,Weber.HADES-collab.2011.JPConfSer,%
Agakichiev.HADES-collab.2010.PLB,Holzmanna.HADES-collab.2010.NPA,Agakichiev.HADES-collab.2009.EPJA,%
Salabura.HADES-collab.2005.NPA,Salabura.HADES-collab.2004.ActaPhysPol,Salabura.HADES-collab.2004.PPNP}.
For the proton-nucleus scattering at energy of proton beam $E_{\rm p}$ below 10~GeV,
experimental data
on dilepton productions were obtained
for \isotope[]{Be} at $E_{\rm p}=1.0$~GeV, 2.1~GeV and 4.9~GeV (DLS, Ref.~\cite{Naudet.1989.PRL}),
for \isotope[]{Be} at $E_{\rm p}=4.9$~GeV (DLS, Ref.~\cite{Roche.1988.PRL}),
for \isotope[]{Nb} at $E_{\rm p}=3.5$~GeV (HADES, Ref.~\cite{Agakishiev.HADES-collab.2012.plb,Weber.HADES-collab.2011.JPConfSer}).

\section{Model
\label{sec.4}}

\subsection{$S$-matrix of $2^{\rm nd}$-order for production of electron-positron pair
\label{sec.4.2}}

The simplest interactions of fermions (electrons, nucleons) and photons are described by S-matrix of the $2^{\rm nd}$ order.
There is only 6 topologically different processes of the second order~\cite{Ahiezer.1981} [see Fig.~3.3 in that book, p.~147].
According to Eq.~(3.2.15) in Ref.~\cite{Ahiezer.1981},
all these processes are described by S-matrix of the second order.%
\footnote{To be closer to the strict formalism in Ref.~\cite{Ahiezer.1981},
formalism in Sect.~\ref{sec.4}
is given in Euclidean metric of spacetime,
see p.~30--32 in that book for connection between formalisms in Euclidean and Pseudo-Euclidean metrics.}
According to Eq.~(3.2.15) in Ref.~\cite{Ahiezer.1981} [see p.~142, 222 in that book], we write down for production of electron-positron pair as
\begin{equation}
\begin{array}{llll}
  S^{(2)} =
  \displaystyle\frac{e^{2}}{2}
  \displaystyle\int
      \mathcal{N} \Bigl(
          \bigl[ \bar{\psi} (x_{1})\, \gamma_{\mu} \, \psi(x_{1}) \bigr]_{\rm n}
          \bigl[ \bar{\psi} (x_{2})\, \gamma_{\nu}\, \psi(x_{2}) \bigr]_{\rm e}
      \Bigr)\;
    A_{\mu}^{a} (x_{1})\, A_{\nu}^{a} (x_{2})\; d^{4}x_{1}\: d^{4}x_{2},
\end{array}
\label{eq.4.2.1}
\end{equation}
where
$\psi(x_{i})$ is operator wave functions for nucleon or electron ($i = 1,2$),
$\bar{\psi}(x_{i}) = \psi^{+}(x_{i})\, \gamma_{4}$,
$A (x_{j})$ is operator wave function of electromagnetic field ($j = 1,2$)
[see Eq.~(2.3.1) in p.~93, Ref.~\cite{Ahiezer.1981}].
Symbol $\mathcal N$ denotes the normal-order product (or normal product) of operators
[for example, see Eq.~(2.3.21) in Ref.~\cite{Ahiezer.1981}, p.~99].
In this formula we take into account rule for connection between operators for photons $A_{\mu}^{a} (x)\, A_{\nu}^{a} (x')$:
If two operators for photons are connected, such operators can be placed close to each other, and their connection is $c$-number
[see Eqs.~(2.3.25), (2.3.26), p.~100, 143 in Ref.~\cite{Ahiezer.1981}].

\subsection{Normal multiplication of operators for fermions and matrix elements
\label{sec.4.3}}

According to Eq.~(2.5.1) in Ref.~\cite{Ahiezer.1981}, 
solution $\psi(x)$ of Dirac equation and its conjugated solution $\bar{\psi}(x)$ are
expanded over
eigenfunctions for fermion $\psi_{s}^{+} (x)$ and $\psi_{r}^{-} (x)$ with positive and negative frequencies as
\begin{equation}
\begin{array}{llllll}
  \psi(x) =
  \displaystyle\sum\limits_{s,\, w_s > 0} a_{s} \psi_{s}^{+} (x) +
  \displaystyle\sum\limits_{r,\, w_s < 0} b_{r}^{+} \psi_{r}^{-} (x), &
  \bar{\psi}(x) =
  \displaystyle\sum\limits_{s,\, w_s > 0} a_{s}^{+} \bar{\psi}_{s}^{+} (x) +
  \displaystyle\sum\limits_{r,\, w_s < 0} b_{r} \bar{\psi}_{r}^{-} (x).
\end{array}
\label{eq.4.3.2}
\end{equation}
Here,
$a_{s}$, $b_{r}$ and $a_{s}^{+}$, $b_{r}^{+}$ are operators acting in space of number of particles --- nucleons or electrons and their antiparticles.
$a_{s}$ is operator of annihilation of electron,
$a_{s}^{+}$ is operator of creation of electron,
$b_{r}$ is operator of annihilation of positron,
$b_{r}^{+}$ is operator of creation of positron (similar is for nucleon),
indexes ``$+$'' and ``$-$'' at wave functions denote states with positive and negative frequencies.
We substitute these representations to Eq.~(\ref{eq.4.2.1}).

Operators for nucleon and electron are anticommute between each other
[see, Ref.~\cite{Ahiezer.1981}, Eq.~(2.5.6), p.~116],
and we obtain~\cite{Maydanyuk_Wolf.2024.arxiv}
\begin{equation}
\begin{array}{lllll}
  \mathcal{N}
    \Bigl\{
      \bar{\psi}_{\rm n} (x_{1}) \gamma_{\mu} \psi_{\rm n} (x_{1})
      \bar{\psi}_{\rm e} (x_{2}) \gamma_{\mu} \psi_{\rm e} (x_{2})
    \Bigr\} =
  \mathcal{N}
    \Bigl\{
      \bar{\psi}_{\rm n} (x_{1}) \gamma_{\mu} \psi_{\rm n} (x_{1})
    \Bigr\}
  \mathcal{N}
    \Bigl\{
      \bar{\psi}_{\rm e} (x_{2}) \gamma_{\mu} \psi_{\rm e} (x_{2})
    \Bigr\}.
\end{array}
\label{eq.4.3.4}
\end{equation}
Following to procedure of quantization of the electronic-positronic field
[see p.~114--123, Eqs.~(2.5.10), (2.5.11) in Ref.~\cite{Ahiezer.1981}),
we calculate matrix element of normal multiplication of operators for electron and positron.
%
We use only states with positive frequencies for electron and negative frequencies for positron.
We obtain
\begin{equation}
\begin{array}{llll}
  \bigl\langle f \bigl| S^{(2)} \bigr| i \bigr\rangle =
  \displaystyle\frac{e^{2}}{2}
  \displaystyle\int
  \Bigl\{
      \bar{\psi}_{s'}^{+} (x_{1}) \gamma_{\mu} \psi_{s}^{+} (x_{1})
  \Bigr\}_{\rm n}
  \Bigl\{
      \bar{\psi}_{r'}^{+} (x_{2}) \gamma_{\nu} \psi_{r}^{-} (x_{2})
  \Bigr\}_{\rm e}
  A_{\mu}^{a} (x_{1})\, A_{\nu}^{a} (x_{2})\; d^{4}x_{1}\: d^{4}x_{2}.
\end{array}
\label{eq.4.5.2}
\end{equation}

\subsection{Integration over time variables and Fourier representation of contraction of wave functions of photons
\label{sec.4.6.3}}

Let's write down expansion of vector potential of electromagnetic field over plane waves as
(see Ref.~\cite{Ahiezer.1981}, Eq.~(2.3.1), p.~93)
\begin{equation}
\begin{array}{lllllll}
  A_{\mu}(x) & = &
  \displaystyle\frac{1}{\sqrt{V}}\,
  \displaystyle\sum\limits_{\vb{k}_{\rm ph},\, \lambda}
    \displaystyle\frac{1}{\sqrt{2 w_{\rm ph}}}\;
    e_{\mu}^{(\lambda)}\,
    \Bigl\{
      c_{k \lambda}\, e^{i\, k_{\rm ph}x} +
      c^{+}_{k \lambda}\, e^{-i\, k_{\rm ph}x}
    \Bigr\}, &
  k_{\rm ph}x = \vb{k}_{\rm ph} \vb{r} - E_{\rm ph}t.
\end{array}
\label{eq.4.6.1.1}
\end{equation}
%
Here,
$E_{\rm ph} = \sqrt{m_{\rm ph}^{2} + \vb{k}_{\rm ph}^{2}}$ is energy of virtual photon,
$w_{\rm ph} = E_{\rm ph}$ is frequency of photon,
$e_{\mu}^{(\lambda)}$ are vectors of polarization of photon,
$m_{\rm ph}$ is mass of virtual photon.


Contraction of two wave functions of photons and its Fourier transformation are
[see Eqs.~(2.3.25), (2.3.26), p.~100, 152 in 
Ref.~\cite{Ahiezer.1981}, we use $d=0$]
%
%
%
%
%
\begin{equation}
\begin{array}{llllllll}
  A^{a}_{\mu}(x_{1})\, A^{a}_{\nu}(x_{2}) =
  D_{c\mu\nu}(x_{1} - x_{2}), \quad
  D_{c\, \mu\nu} (x) =
  \displaystyle\frac{1}{(2\, \pi)^{4}}
  \displaystyle\int\limits
    D_{c\, \mu\nu} (k)\, e^{-i\, kx}\; d^{4}k, \\
  D_{c\, \mu\nu} (k) =
  \displaystyle\frac{1}{i\, (k^{2} - i0)}
  \Bigl(
    \delta_{\mu\nu} -
    d\, \displaystyle\frac{k_{\mu}k_{\nu}}{k^{2}}
  \Bigr).
\end{array}
\label{eq.method1.1.5}
\end{equation}
%

Wave function of electron and positron can be written down as
[see Eq.~(2.5.15) in p.~120, Eq.~(4.2.2) in p.~223 in Ref.~\cite{Ahiezer.1981}]
\begin{equation}
\begin{array}{llllllll}
  \psi_{p\, a}^{\rm (e, +)} (x)         = \varphi_{p\, a}^{\rm (e, +)}\, (\vb{r})\, e^{-iE_{\rm e}t}, &
  \varphi_{p\, a}^{\rm (e, +)} (\vb{r}) = \displaystyle\frac{1}{\sqrt{2\,V\, E_{\rm e}}}\; u_{a}\,(+p)\, e^{i\, \vb{p r}}, \\
\vspace{0.5mm}
  \bar{\psi}_{p\, a}^{\rm (e, +)} (x)         = \bar{\varphi}_{p\, a}^{\rm (e, +)} (\vb{r})\, e^{iE_{\rm e}t}, &
  \bar{\varphi}_{p\, a}^{\rm (e, +)} (\vb{r}) = \displaystyle\frac{1}{\sqrt{2\,V\, E_{\rm e}}}\; \bar{u}_{a}\,(p)\, e^{- i\, \vb{p r}}, \\
  \psi_{-p\, a}^{\rm (e, -)} (x)         = \varphi_{-p\, a}^{\rm (e, -)}\, (\vb{r})\, e^{iE_{\rm e}t}, &
  \varphi_{-p\, a}^{\rm (e, -)} (\vb{r}) = \displaystyle\frac{1}{\sqrt{2\,V\, E_{\rm e}}}\; u_{a}\,(-p)\, e^{- i\, \vb{p r}}, \\
  \bar{\psi}_{-p\, a}^{\rm (e, -)} (x)         = \bar{\varphi}_{-p\, a}^{\rm (e, -)} (\vb{r})\, e^{-iE_{\rm e}t}, &
  \bar{\varphi}_{-p\, a}^{\rm (e, -)} (\vb{r}) = \displaystyle\frac{1}{\sqrt{2\,V\, E_{\rm e}}}\; \bar{u}_{a}\,(-p)\, e^{i\, \vb{p r}},
\end{array}
\label{eq.4.6.3.3}
\end{equation}
where $u_{a}$ is bispinor for electron (similar formulas are for nucleon).
%
Substituting wave functions (\ref{eq.4.6.3.3}) and connection (\ref{eq.method1.1.5}) to Eq.~(\ref{eq.4.5.2}), we obtain:
\begin{equation}
\begin{array}{llll}
\vspace{1.0mm}
  \bigl\langle f \bigl| S^{(2)} \bigr| i \bigr\rangle =
  \displaystyle\frac{e^{2}}{2}\,
  \displaystyle\int
    \bar{\varphi}_{s', {\rm n}}^{+} (\vb{r_{1}}) \gamma_{\mu}\, \varphi_{s, {\rm n}}^{+} (\vb{r_{1}}) e^{i\, (E_{\rm n}' - E_{\rm n}) t_{1}}
    \bar{\varphi}_{r', {\rm e}}^{+} (\vb{r_{2}}) \gamma_{\mu}\, \varphi_{r, {\rm e}}^{-} (\vb{r_{2}}) e^{i\, (E_{\rm e} + E_{\rm pos}) t_{2}} \\
  \times \quad
  \displaystyle\frac{1}{(2\, \pi)^{4}}
  \displaystyle\int\limits
    D_{c} (k)\, e^{- i\, k(x_{1} - x_{2})}\; d^{4}k\;
    d^{4}x_{1}\, d^{4}x_{2}, \quad
  D_{c} (k) \equiv D_{c\, \mu\mu} (k) \Bigr|_{d=0} =
    \displaystyle\frac{1}{i\, (k^{2} - i0)} .
\end{array}
\label{eq.method1.1.6}
\end{equation}

We calculate integration in this matrix element over
time variables $t_{1}$, $t_{2}$,
space variables $\vb{r}_{1}$, $\vb{r}_{2}$ of nucleons and electron,
and 4-momentum $k_{0}$, $\vb{k}$ of virtual photon.
Calculations are straightforward
(see Appendix~C in Ref.~\cite{Maydanyuk_Wolf.2024.arxiv}) and we obtain:
\begin{equation}
\begin{array}{llllll}
  \bigl\langle f \bigl| S^{(2)} \bigr| i \bigr\rangle =
  \displaystyle\frac{2 \pi w\: M_{\rm n}^{(\mu)} M_{\rm e}^{(\mu)}} {i\, (k_{{\rm ph}, 0}^{2} - \vb{k}_{\rm ph}^{2})}\,
  \delta (k_{\rm n}' - k_{\rm n} + k_{\rm e} + k_{\rm pos}), &

  E_{\rm ph} = E_{\rm e} + E_{\rm pos},\:
  \vb{k_{\rm ph}} = \vb{k_{\rm e}} + \vb{k_{\rm pos}}.
\end{array}
\label{eq.method1.1.7}
\end{equation}
%

\subsection{Non-relativistic formalism of many-nucleon system in the proton-nucleus scattering
\label{sec.method1.10}}

Let us formulate the model for production of dileptons in the scattering of protons on nuclei.
At first, we write down matrix element of emission of photons for many-nucleon system without magnetic moments of nucleons%
~\cite{Maydanyuk.2023.PRC.delta}
%
%
%
\begin{equation}
\begin{array}{llllllll}
  \langle \Psi_{f} |\, \hat{H}_{\gamma} |\, \Psi_{i} \rangle \;\; = \;\;
  \Bigl\langle \Psi_{f} \Bigl|\,
  \displaystyle\sum\limits_{i=1}^{A}
  \Bigl\{
    i\hbar\, \displaystyle\frac{z_{i}\, e}{m_{\rm p}c}\,
    \displaystyle\frac{1}{\sqrt{2\, w_{\rm ph}}}\,
    \vb{\displaystyle\frac{d}{dr_{i} }}\,
    e^{- i\, \vb{k_{\rm ph} r_{i}}}
  \Bigr\}
  \Bigl|\, \Psi_{i} \Bigr\rangle.
\end{array}
\label{eq.method1.10.2}
\end{equation}
Here we already use many-nucleon wave functions of proton-nucleus scattering,
$\Psi_{i}$ and $\Psi_{f}$, in states before emission of photon ($i$-state) and after such an emission ($f$-state).
%
%
We obtain analogy in construction of matrix element of emission of photon
between formalisms for one nucleon moving in external field and many-nucleon system in the nuclear scattering.
Calculation of such matrix elements of emission of photon (\ref{eq.method1.10.2}) are given in Ref.~\cite{Maydanyuk.2023.PRC.delta}
(this is result of researches
\cite{Maydanyuk.2012.PRC,Maydanyuk_Zhang.2015.PRC,Maydanyuk_Zhang_Zou.2016.PRC,%
Maydanyuk_Vasilevsky.2023.PRC,Shaulskyi_Maydanyuk_Vasilevsky.2024.PRC})
%
\begin{equation}
\begin{array}{llllll}
\vspace{0.5mm}
 \vb{M}_{A,\, \mu} & = &
  \langle \Psi_{f} |\, \hat{H}_{\gamma} |\, \Psi_{i} \rangle \;\; = \;\;
  \sqrt{\displaystyle\frac{2\pi\, c^{2}}{\hbar w_{\rm ph}}}\,
  \vb{M}_{\rm full}, \quad

  M_{A}^{(\alpha)} =
 \vb{M}_{A,\, \mu}\, \vb{e}^{(\alpha)}_{\mu} =
 f_{A} \cdot \delta_{\alpha\, 2}, \\
  f_{A} & = &
  -\, \sqrt{\displaystyle\frac{2\pi\, c^{2}}{3\, \hbar w_{\rm ph}}}\, \cdot
  \hbar\, (2\pi)^{3} \displaystyle\frac{\mu_{N}\,  m_{\rm p}}{\mu}\;
  Z_{\rm eff}^{\rm (mon,\, 0)} \cdot
  \Bigl\{
    J_{1}(0,1,0) -
    \displaystyle\frac{47}{40} \sqrt{\displaystyle\frac{1}{2}} \cdot J_{1}(0,1,2)
  \Bigr\},
\end{array}
\label{eq.6-nucleus.1.2}
\end{equation}
where
$\mu_{N} = e\hbar / (2m_{\rm p}c)$ is nuclear magneton,
$\mu = m_{\rm p} m_{A} / (m_{\rm p} + m_{A})$ is reduced mass of proton and nucleus,
$m_{\rm p}$ and $m_{A}$ are masses of proton and nucleus,
$Z_{\rm eff}^{\rm (mon,\, 0)}$ is effective electric charge of proton-nucleus system
defined in Eq.~(25) in Ref.~\cite{Maydanyuk.2023.PRC.delta},
%
%
$J_{1}(0,1,0)$ and $J_{1}(0,1,2)$ are the radial integrals defined as
($r$ is relative distance between the scattered proton and center-of-mass of nucleus-target)
%
\begin{equation}
\begin{array}{llllll}
  J_{1}^{real} (l_{i},l_{f},n) & = &
    \displaystyle\int\limits^{+\infty}_{0} \displaystyle\frac{dR_{i}(r, l_{i})}{dr}\: R^{*}_{f}(l_{f},r)\, j_{n}(k_{\rm ph}r)\; r^{2} dr.
\end{array}
\label{eq.resultingformula.6}
\end{equation}
%

Tensor associated with production of electron-positron pair has starnard formalism~\cite{Maydanyuk_Wolf.2024.arxiv}.
%
%
Square of $O$-matrix element for production of lepton pair is
matrix multiplication of dileptonic matrix elements 
and nuclear matric elements
~\cite{Maydanyuk_Wolf.2024.arxiv}
%
\begin{equation}
\begin{array}{rllll}
  \Bigl| \langle f | O^{(2)} | i \rangle \Bigr|^{2}
  \displaystyle\frac{d^{3}k_{\rm e}}{(2\pi)^{3}}
  \displaystyle\frac{d^{3}k_{\rm pos}}{(2\pi)^{3}} =
  \displaystyle\frac{|f_{A}|^{2}\; e^{2} w^{2}\, |\vb{k}_{\rm e}|} {4\, (k_{{\rm ph},\, 0}^{2} - \vb{k}_{\rm ph}^{2})^{2}}\,
  \Bigl[ 1 - 2 \sin^{2} (c_{\theta} \theta) - \cos{2 \theta} \Bigr]
  \displaystyle\frac{1 + \cos{2\, \theta}}{1 - \cos{2 \theta}}\: do_{\rm e}.
\end{array}
\label{eq.6-nucleus.2.3}
\end{equation}
%
%
Here, $\theta$ is angle between direction of virtual photon emission (defined as $\vb{k}_{\rm ph}/|\vb{k}_{\rm ph}|$) and
direction of electron emission (defined as $\vb{k}_{\rm e} / |\vb{k}_{\rm e}|$).


\subsection{Inclusion of incoherent processes 
\label{sec.model.incoh}}

Formalism above is developed on the basis of coherent emission of virtual photon of electric type $M_{p}^{(E,\, {\rm mon},\, 0)}$.
So, we have to include also other terms of emission of photons.
Such a formalism was developed in Ref.~\cite{Maydanyuk.2023.PRC.delta}.
According to this approach, instead of (\ref{eq.6-nucleus.1.2}) full nuclear matrix element of emission of photon is written down as
%
%
%
%
%
\begin{equation}
\begin{array}{lllll}
  M_{\rm full} =
  M_{P} +
  M_{p}^{(E,\, {\rm mon},\, 0)} + M_{p}^{(M,\, {\rm mon},\, 0)} + M_{\Delta M} + M_{k}, &
\end{array}
\label{eq.model.incoh.1.5}
\end{equation}
%
%
\begin{equation}
\begin{array}{lll}
\vspace{1.0mm}
  M_{p}^{(E,\, {\rm mon},\, 0)} =
  -\, \hbar\, (2\pi)^{3}\,
  \displaystyle\frac{\mu_{N}}{\sqrt{3}}\,
  \displaystyle\frac{m_{\rm p}\, Z_{\rm eff}^{\rm (mon,\, 0)}}{\mu}\,
  \Bigl(
    J_{1}(0,1,0) -
    \displaystyle\frac{47}{40} \sqrt{\displaystyle\frac{1}{2}}\, J_{1}(0,1,2)
  \Bigr), \\
\vspace{1.0mm}
  M_{p}^{(E,\, {\rm mon},\, 0)} + M_{p}^{(M,\, {\rm mon},\, 0)} =
%
  M_{p}^{(E,\, {\rm mon},\, 0)}\,
  \Bigl( 1 + i\: \displaystyle\frac{\mu^{2}} {2\, m_{\rm p}^{2}\, Z_{\rm eff}^{\rm (mon,\, 0)}}\; \alpha \Bigr), \\
\vspace{1.0mm}
  M_{\Delta M} =
  \hbar\, (2\pi)^{3}\,
  \displaystyle\frac{\sqrt{3}}{2}\,
  \mu_{N}\, k_{\rm ph}\,
  f_{A} \cdot Z_{\rm A} (\vb{k}_{\rm ph}) \cdot \tilde{J}\, (- c_{\rm p}, 0,1,1), \\
  M_{\Delta M} + M_{k} =
  - \hbar (2\pi)^{3}
  \displaystyle\frac{\sqrt{3}}{2}
  \mu_{N} k_{\rm ph}
  \Bigl\{
    \displaystyle\frac{A+1}{2A} \bar{\mu}_{\rm pn}^{(A)} Z_{\rm A} (\vb{k}_{\rm ph}) \tilde{J} (- c_{\rm p}, 0,1,1) +
    \mu_{\rm p} z_{\rm p} \tilde{J} (c_{A}, 0,1,1)
  \Bigr\}.
\end{array}
\label{eq.model.incoh.1.6}
\end{equation}
Here,
$M_{p}^{(M,\, {\rm mon},\, 0)}$ is coherent matrix element of magnetic type,
$M_{\Delta M}$ and $M_{k}$ are incoherent matrix element of magnetic type and matrix element of background.
$\bar{\mu}_{\rm pn}^{\rm (A)} = \mu_{\rm p}^{\rm (an)} + \kappa_{A}\,\mu_{\rm n}^{\rm (an)}$,
$\mu_{\rm p}^{\rm (an)} = 2.79284734462$ is magnetic moment for proton,
$\mu_{\rm n}^{\rm (an)} = -1.91304273$ is magnetic moment for neutron
(measured in units of nuclear magneton $\mu_{N}$, see Ref.~\cite{RewPartPhys_PDG.2018}),
$\kappa_{A} = N_{A}/Z_{A}$,
$A$ and $N_{A}$ are numbers of nucleons and neutrons in nucleus with index $A$.
%
Parameter $\alpha$ and
effective electric charge are~\cite{Maydanyuk.2023.PRC.delta}
%
\begin{equation}
\begin{array}{llllll}
  \alpha =
  \displaystyle\frac{m_{\rm p}}{\mu}\;
  \Bigl[
    \displaystyle\frac{m_{\rm p}}{m_{A}}\, Z_{\rm A} (\vb{k}_{\rm ph})\, \bar{\mu}_{\rm pn}^{\rm (A)} -
    z_{\rm p}\, \mu_{\rm p}
  \Bigr], &

  Z_{\rm eff}^{\rm (mon)} (\vb{k}_{\rm ph}) = \displaystyle\frac{m_{\rm p}\, Z_{A}(\vb{k}_{\rm ph}) - m_{A}\, z_{\rm p}}{m_{A} + m_{\rm p}}.
%
%
\end{array}
\label{eq.app.2.resultingformulas.6}
\end{equation}
%
%
%
$Z_{A}(\vb{k}_{\rm ph})$ is electric form factor of nucleus-target describing structure of this nucleus.
This form factor and all other form factors in this formalism are defined in
%
%
Sect.~I in Supplemental Material in Ref.~\cite{Maydanyuk.2023.PRC.delta} [see Eqs. (16), (21), (23), (25) there].


%

\subsection{Cross section of production of leptons pair
\label{sec.model.bremprobability}}

Cross sections of the production of leptons pairs can be defined on the basis of logic of determination of
cross section of emission of bremsstrahlung photons in nucleon-nucleus and nucleus-nucleus scattering~\cite{Maydanyuk.2023.PRC.delta,Maydanyuk_Zhang_Zou.2016.PRC,Maydanyuk.2012.PRC,Maydanyuk_Zhang.2015.PRC}.
We define cross section for production of electron positron pair as%
~\cite{Maydanyuk_Wolf.2024.arxiv}
\begin{equation}
\begin{array}{lllllll}
  d^{2} \sigma^{\rm (lep)} = &
  \displaystyle\frac{\pi^{2} e^{2}}{2\, c^{5}}
  \displaystyle\frac{|f_{A} (M, \chi)|^{2} w^{2}} {(k_{{\rm ph},\, 0}^{2} - \vb{k}_{\rm ph}^{2})^{2}}\:
  \displaystyle\frac{E_{\rm p}} {k_{\rm p}\, E_{\rm e}^{2}}\,
  |\vb{k}_{\rm e}|\,
  \Bigl[ 1 - 2\, \sin^{2} (c_{\theta}\, \theta) - \cos{2 \theta} \Bigr]\;
  \displaystyle\frac{1 + \cos{2\, \theta}}{1 - \cos{2\, \theta}}\; do_{\rm e}.
%
\end{array}
\label{eq.model.crosssection.2.9}
\end{equation}
%
%
Here, $|\vb{k}_{\rm e}| = M / 2$ is taken from Ref.~\cite{Zetenyi_Wolf.2012.PRC} neglecting the electron mass.
$E_{\rm p}$ is kinetic energy of relative motion in the center-of-mass frame between proton in beam and nucleus-target, 
$k_{\rm p} = \sqrt{2E\mu} / \hbar$,
$\mu = m_{A}\, m_{\rm p}/(m_{A} + m_{\rm p})$ is reduced mass.
$M = \sqrt{k_{\rm ph}^{2}}$ is the dilepton invariant mass~\cite{Zetenyi_Wolf.2012.PRC}
($k_{\rm ph}$ is 4-momentum of virtual photon).
In this paper the matrix elements are calculated on the basis of wave functions with quantum numbers $l_{i}=0$, $l_{f}=1$ and $l_{\rm ph}=1$
[here, $l_{i}$ and $l_{f}$ are orbital quantum numbers of wave function $\Phi_{\rm p - nucl} (\vb{r})$
of relative motion in scattering between proton and nucleus
in states before emission of photon and after this emission,
$l_{\rm ph}$ is orbital quantum number of photon in the multipole approach].
%
%
All components of virtual photon were studied in Ref.~\cite{Maydanyuk_Wolf.2024.arxiv}
and we omit this analysis in this paper.

\section{Analisys
\label{sec.analysis}}

\subsection{The coherent and incoherent processes in production of dilepton pair in the scattering $p + \isotope[9]{Be}$ and experimental data
\label{sec.analysis.1}}

\begin{figure}[htbp]
\centerline{\includegraphics[width=74mm]{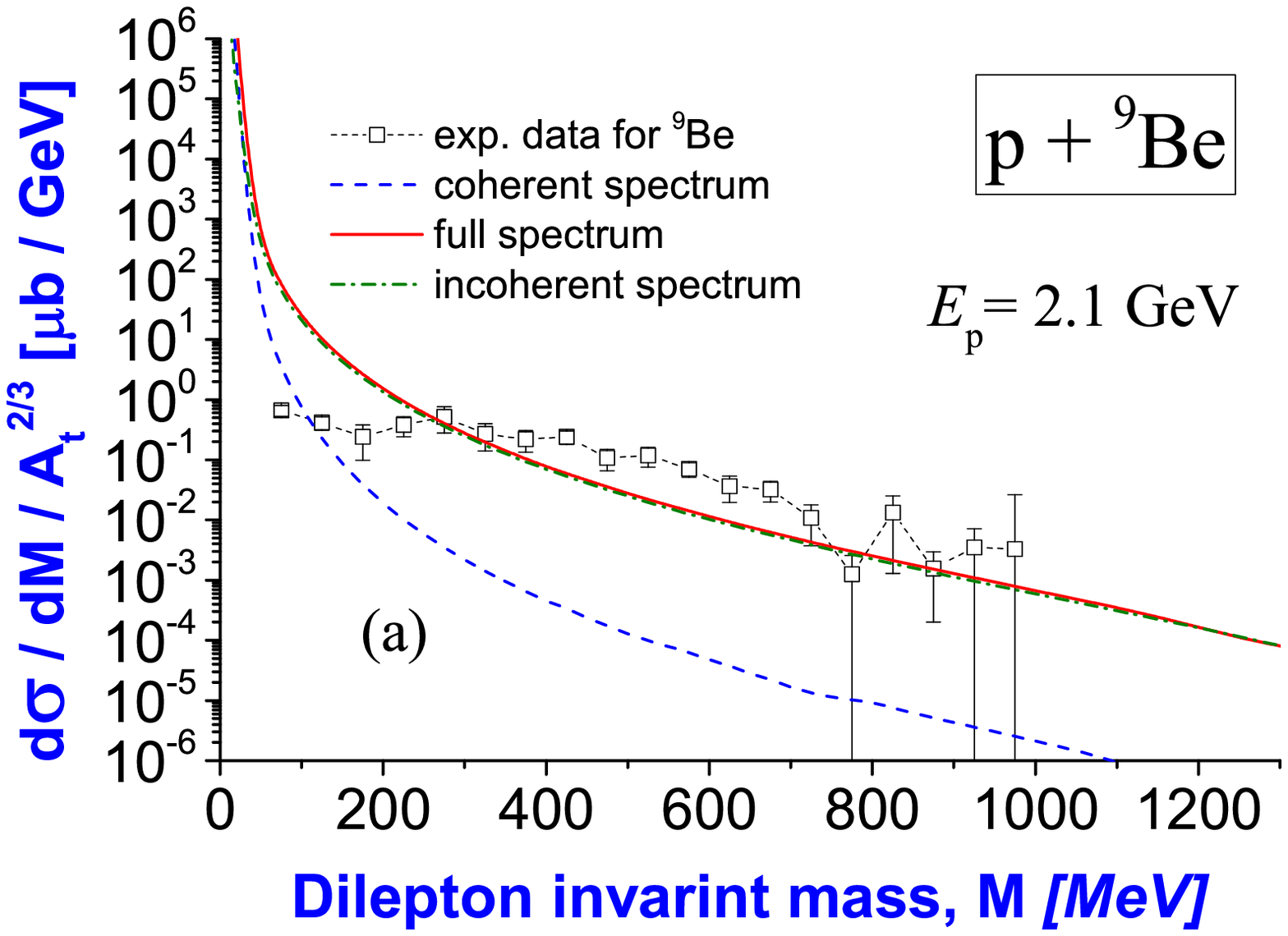}
\hspace{-4mm}\includegraphics[width=74mm]{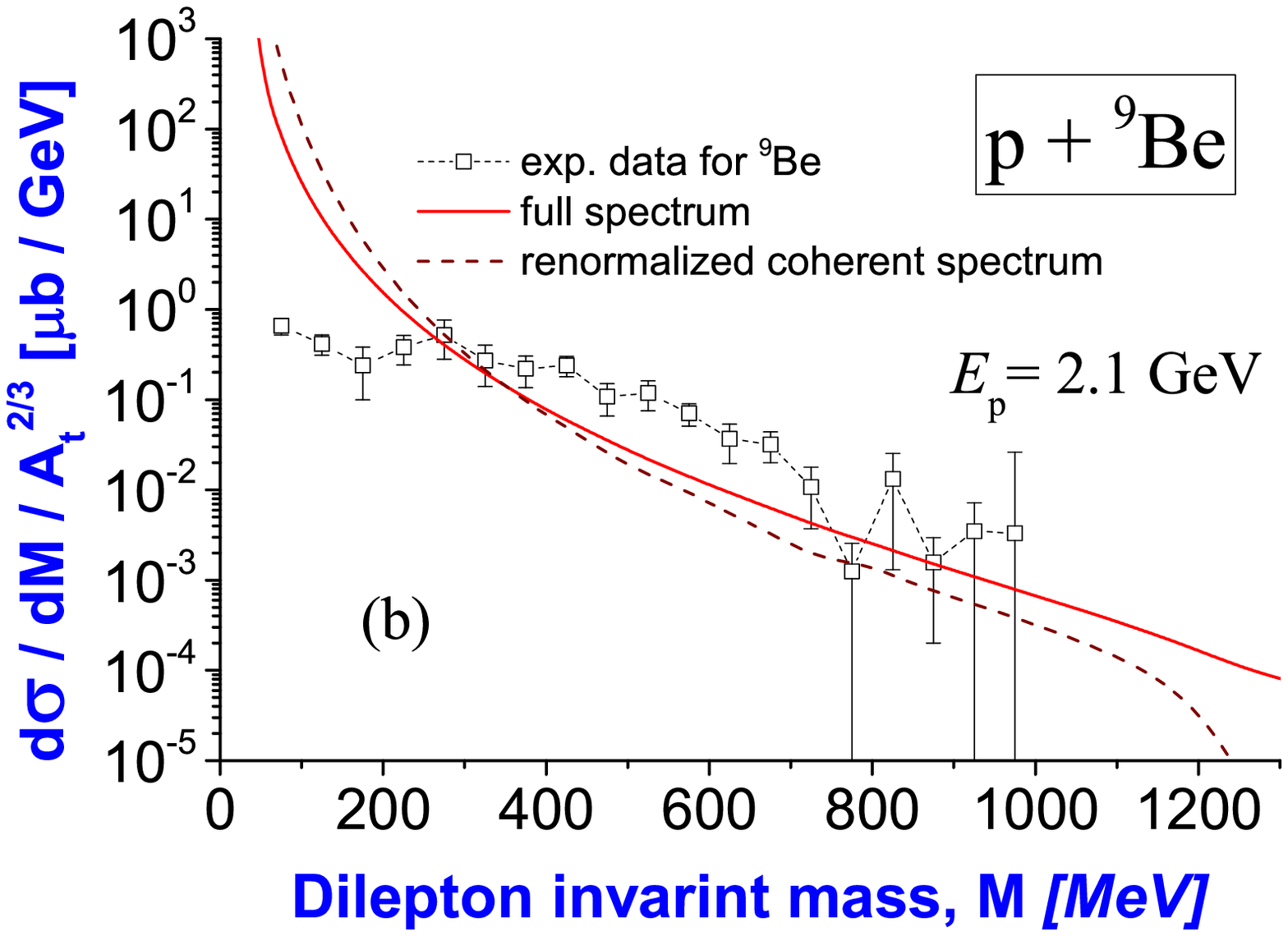}}
\vspace{-4mm}
\caption{\small (Color online)
The calculated full cross section and different contributions for production of electron-positron pair
in the scattering of protons off the \isotope[9]{Be} nuclei at energy of proton beam of $E_{\rm p}=2.1$~GeV
in comparison with experimental data~\cite{Naudet.1989.PRL}
[
$Z_{A} (\vb{k}) = z_{A}$,
$z_{A}$ is electric charge of nucleus,
the calculated full spectrum is normalized on experimental data]
Here, experimental data given by open rectangles are extracted from Ref.~\cite{Naudet.1989.PRL}.
Panel (a):
Calculated full cross section, coherent and incoherent contributions in comparison with experimental data.
Panel (b):
The calculated full cross section in comparison with the coherent contribution after its additional re-normalization on experimental data.
\label{fig.1}}
\end{figure}

We will analyze experimental data of production of dileptons in scattering of protons off nuclei.
Theses are data obtained by DLS Collaboration for \isotope[]{Be} at proton beam energies of
1.0~GeV, 2.1~GeV and 4~GeV~\cite{Naudet.1989.PRL} and
by HADES Collaboration for \isotope[]{Nb} at proton beam energies of 3.5~GeV~\cite{Agakishiev.HADES-collab.2012.plb,Weber.HADES-collab.2011.JPConfSer}.
Note that those data were analyzed in details in different approaches (see Refs.~\cite{Wolf.1993.PPNP,Wolf.1990.NPA}).
We calculate wave function of relative motion between proton and center-of-mass of nucleus numerically
concerning to the proton-nucleus potential in form of $V (r) = v_{c}(r) + v_{N}(r) + v_{\rm so}(r) + v_{l} (r)$,
where $v_{c}(r)$, $v_{N}(r)$, $v_{\rm so}(r)$, and $v_{l} (r)$ are Coulomb, nuclear, spin-orbital, and centrifugal components, respectively.
Nuclear parameters are defined in Eqs.~(46)--(47) in Ref.~\cite{Maydanyuk_Zhang.2015.PRC}.
%
Calculations of the full cross section and contributions for \isotope[9]{Be} in comparison with experimental data are shown in Fig.~\ref{fig.1}.
%
Our model provides the spectrum in good agreement with experimental data (with the exception of low part of energy region).
One can see that incoherent processes are larger than coherent ones.

\subsection{Are dilepton pairs produced from nucleus-target during resonant scattering
or in external region outside this nucleus during potential scattering?
\label{sec.analysis.2}}

Let us clarify the next question, in which space region are electron-positron pairs produced?
Is this should be internal space region of nucleus (process of resonant scattering of proton on nucleus \isotope[9]{Be})?
Or is this should be external region outside nucleus \isotope[9]{Be} and
dileptons are mainly produced during potential scattering?

\begin{figure}[htbp]
\centerline{\includegraphics[width=74mm]{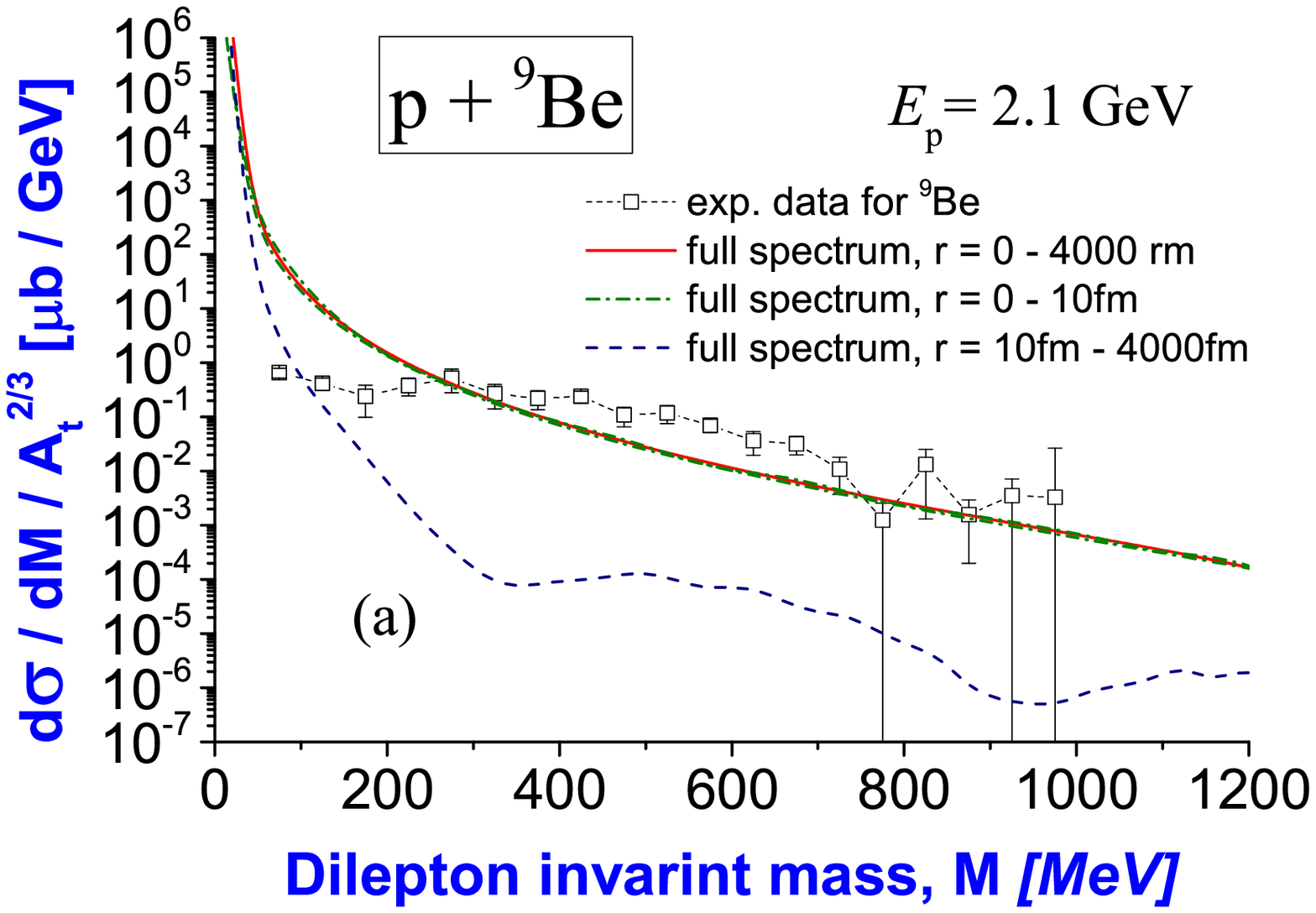}
\hspace{-4mm}\includegraphics[width=74mm]{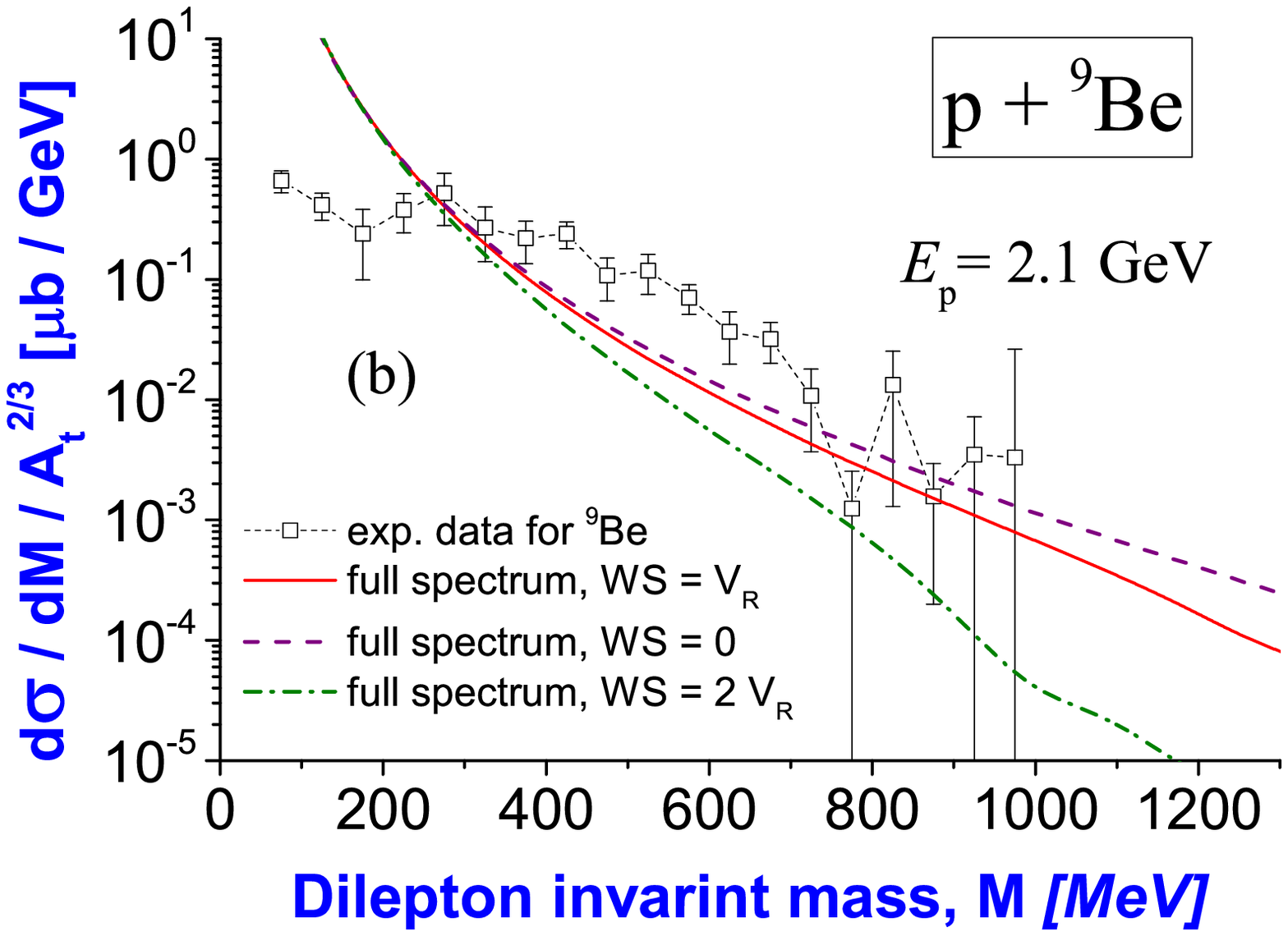}}
\vspace{-5mm}
\caption{\small 
The calculated full cross section of dileptons in $p + \isotope[9]{Be}$ at $E_{\rm p}=2.1$~GeV
in comparison with experimental data~\cite{Naudet.1989.PRL} 
[$Z_{A} (\vb{k}) = z_{A}$,
$z_{A}$ is electric charge of nucleus].
Open rectangles are experimental data~\cite{Naudet.1989.PRL}.
Panel (a):
Comparison between
contribution of dileptons produced in region outside nucleus (potential scattering),
and contribution of dileptons produced in region of nucleus during scattering (resonant scattering).
Panel (b):
The calculated cross sections at different strengths $WS$ of nuclear part of proton-nucleus potential
(see Eqs.~(46)--(47) in Ref.~\cite{Maydanyuk_Zhang.2015.PRC} for definitions).
%
\label{fig.2}}
\end{figure}

Our model is constructed in space representation.
In integration we choose space region with minimal and maximal boundaries, given by values $R_{\rm min}$ and $R_{\rm max}$,
and calculate integrals (\ref{eq.resultingformula.6}).
%
%
%
We calculate contributions of dileptons produced
in the external region during scattering outside nucleus-target
and in the internal region of nucleus-target during scattering.
Results of such calculations for $p + \isotope[9]{Be}$ 
are shown in Fig.~\ref{fig.2}~(a).
%
One can see that production from external region outside nucleus is smaller than from internal region of nucleus.
From result in Fig.~\ref{fig.2}~(a) opinion can be appeared that role of nuclear forces in production of dileptons at such energies is not important.
But if potential scattering is very small,
then one can conclude that mechanisms of scattering are not important in this problem.
But, let us analyze calculations with and without inclusion of nuclear part of potential of interactions between proton and nucleus in scattering.
We calculate cross section of dileptons where we remove nuclear part of proton-nucleus potential.
Comparing these new calculations shown in Fig.~\ref{fig.2}~(b) with previous results in Fig.~\ref{fig.2}~(a),
one can see that our model provides different cross sections at different strengths $V_{R}$ of nuclear part of potential.





\subsection{Can structure of nucleus-target influence on production of electron-positron pairs
in proton-nucleus scattering
\label{sec.analysis.3}}

\begin{figure}[htbp]
\centerline{\includegraphics[width=80mm]{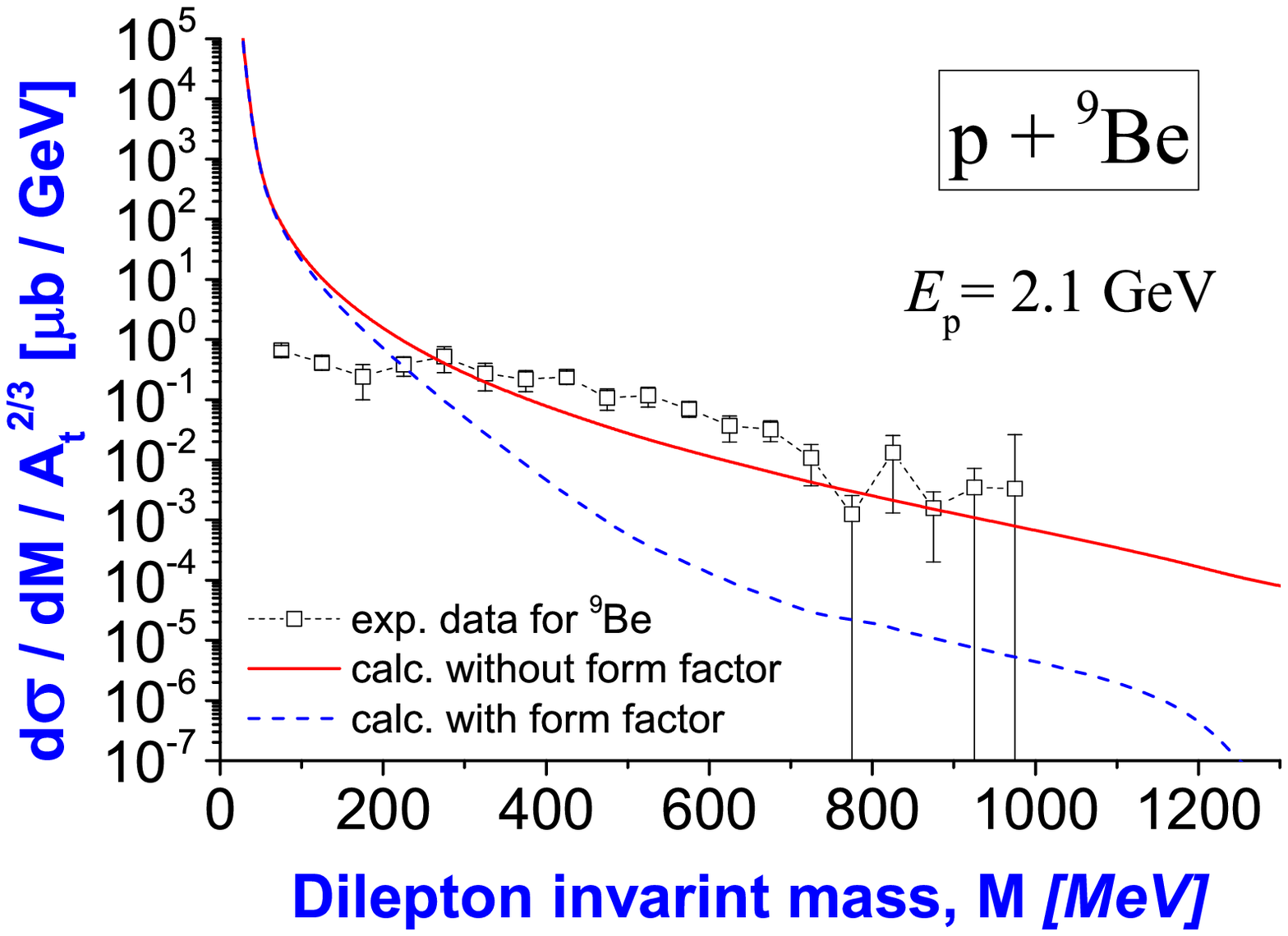}}
\vspace{-4.5mm}
\caption{\small 
The calculated full cross sections of production of electron-positron pair
with included structure of nucleus (see dashed blue line) and
without such structure (see red solid line)
in the scattering of protons off the \isotope[9]{Be} nuclei
\label{fig.3}}
\end{figure}

There is popular point of view that structure of nuclei does not influence much of study of production of dileptons at collision of nuclei at the studied energies.
However, strict quantum mechanical formalism shows that this structure of nuclei is clearly visible in spectra of dileptons.
Indication on this phenomenon we found when we calculated emission of bremsstrahlung photons in scattering of light nuclei~\cite{Maydanyuk_Vasilevsky.2023.PRC}.
We found that even at higher energies of beam structure of these nuclei is more visible in bremsstrahlung spectra than for more low energies
[see Fig.~5~(b) in that paper].
This can be understood from Eq.~(61) in Ref.~\cite{Maydanyuk_Vasilevsky.2023.PRC}.
%
Structure of nucleus is described by its electric form factor in our formalism~\cite{Maydanyuk_Vasilevsky.2023.PRC}
\begin{equation}
\begin{array}{llllllll}
  Z_{A} (\vb{k}) \;\; = \;\;
  z_{A}\;
  \exp\Bigl\{
    -\, \displaystyle\frac{A - 1}{4\; A}\;
    (\vb{k}_{\rm ph},\,\vb{b})^{2}
  \Bigr\},
\end{array}
\label{eq.analysis.3.1}
\end{equation}
where
$z_{A}$ electric charge of nucleus,
$b$ is the oscillator length.
%
Difference between spectra with structure of nucleus and without it is essential (see Fig.~\ref{fig.3}).
%
%

\section{Conclusions and perspective
\label{sec.conclusions}}

On the basis of new model with basis of previous formalism~%
\cite{Maydanyuk_Wolf.2024.arxiv,%
Maydanyuk_Vasilevsky.2023.PRC,Shaulskyi_Maydanyuk_Vasilevsky.2024.PRC,%
Maydanyuk_Zhang.2015.PRC,Maydanyuk.2012.PRC,Maydanyuk_Zhang_Zou.2016.PRC,%
Maydanyuk.2023.PRC.delta},
production of dileptons in the scattering of protons off nuclei is studied
with focus on
aspects of scattering and structure of nuclei.
%
We find the following.

\begin{itemize}
\item
Tendency of calculated full cross section of dileptons production for $p + \isotope[9]{B}$ at $E_{\rm p} = 2.1$~GeV
is in good agreement (with exception of low energies of invariant masses)
with experimental data~\cite{Naudet.1989.PRL} 
[see Fig.~\ref{fig.1}~(a)].

\item
Incoherent contribution is larger than the coherent one [see Fig.~\ref{fig.1}].

\item
Dileptons are mainly produced from space region of nucleus-target during scattering [see Fig.~\ref{fig.2}~(a)].
Our model allows to extract information about nuclear parameters of proton-nucleus potential
[see Fig.~\ref{fig.2}~(b)].


\item
Influence of structure of nucleus on calculations of cross sections of production of electron positron pairs is essential
(see Fig.~\ref{fig.3}).

\end{itemize}

\section*{Acknowledgements
\label{sec.acknowledgements}}

Authors are thanks to Profs. Pengming Zhang, Liping Zou, V.~S.~Vasilevsky, A.~G.~Magner for
for many fruitful discussions concerning to photon emission in nuclear reactions, cluster models.
Authors thank the support of OTKA grant K138277.

\end{document}